\documentclass[reprint,aip,jcp,showpacs,superscriptaddress,preprintnumbers,amsmath,amssymb,floatfix]{revtex4-1}
\usepackage{graphicx}
\usepackage{tabularx}
\usepackage{dcolumn}
\usepackage{amsmath}
\usepackage{amssymb}
\usepackage{bm}
\usepackage{graphics}
\usepackage{subfigure}
\usepackage{booktabs}
\usepackage{array}
\usepackage{xcolor}
\begin{document}

\newcommand{\arhplus}{\ensuremath{\mathrm ArH^+}}
\newcommand{\htwo}{\ensuremath{\mathrm H_2}}
\newcommand{\htwoplus}{\ensuremath{\mathrm H_2^+}}
\newcommand{\hhh}{\ensuremath{\mathrm H_3^+}}
\newcommand{\cm}{cm$^{-1}$}

\title{Tunnelling splitting in the phosphine molecule}

\author{Clara Sousa-Silva}%
\author{Jonathan Tennyson}%
\author{Sergey N. Yurchenko}%

\affiliation{Department of Physics and Astronomy, University College London,
London WC1E 6BT, UK}%

\date{\today}

\begin{abstract}

Splitting due to tunnelling via the potential energy barrier has
  played a significant role in the study of molecular spectra since
  the early days of spectroscopy. The observation of the ammonia
  doublet led to attempts to find a phosphine analogous, but these
  have so far failed due to its considerably higher barrier.
Full dimensional, variational nuclear motion calculations
are used to predict splittings as a function of excitation energy.
Simulated spectra suggest that such splittings should be observable
in the near infrared via overtones of the $\nu_2$ bending mode starting with $4\nu_2$.
\end{abstract}

\maketitle

The umbrella mode in ammonia provides a textbook example of tunnelling
splitting.\cite{80Bell} That the inversion of pyramidal NH$_3$ should
lead to an observable splitting of the energy levels was first
theoretically predicted in 1932 \cite{32DeUh} and then detected using
microwave spectroscopy in 1934.\cite{34ClWiXX} The subtle effects of
this tunnelling on the energy levels of ammonia are now
well-studied.\cite{16CsFuxx.NH3}
As a direct analogue of ammonia, phosphine can also be expected to
display splitting of its energy levels due to the tunnelling effect.
However, splitting in PH$_3$ is yet to be observed, despite multiple
attempts spread over more than 80 years.\cite{33WrRa, 53StOeBe,
  69HeGo, 71DaNeWo, 73MaSaOl, 78SpPa, 81BeBuGe} Although otherwise similar to
ammonia, phosphine has a larger mass and a much
higher and wider barrier which makes for a much smaller splitting of
the energy levels \cite{80Bell}. {\it Ab initio}
calculations of the energy barrier
for phosphine range from 12~270 cm$^{-1}$ to 12~560 cm$^{-1}$,\cite{92ScLaPy.PH3}
while the only empirical estimate gave a slightly
lower value of 11~030  cm$^{-1}$.\cite{54We}

%
%

Experimentally, the infrared spectrum of PH$_3$ has been well studied
(see Table 1 of Sousa-Silva {\it et al.}\cite{jt556} and the recent
study by Devi {\it et al.}\cite{14DeKlSa.PH3}). While most of this
work has concentrated on the region below 3500 \cm, where our
calculations suggest the tunnelling splitting is very small, Ulenikov
{\it et al.}\cite{02UlBeKo.PH3} report observed spectra between 1750
and 9200 \cm~and clearly demonstrate that PH$_3$ spectra can be observed
at higher frequencies.

\begin{figure}
\centering
\includegraphics[width=0.48\textwidth]{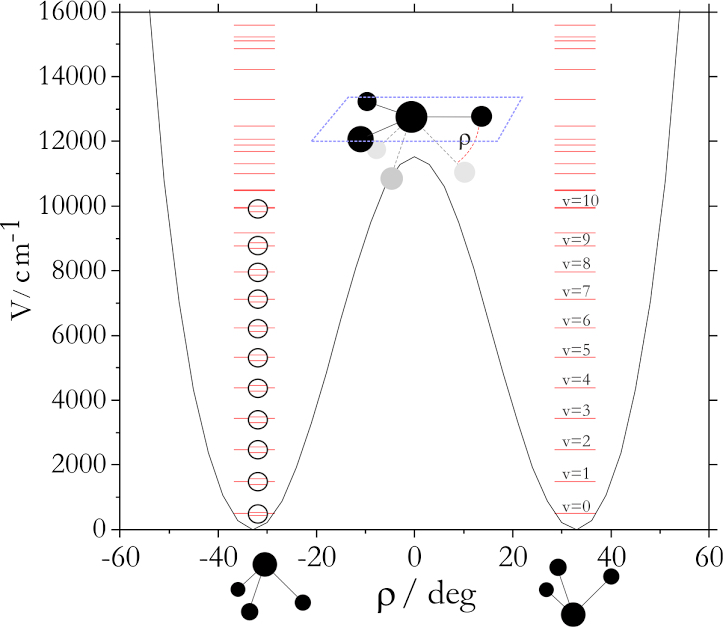}
\caption{\small{Splitting of the energy levels for phosphine, showing the splitting for the ground state and the vibrational excitations up to $v=10$ in the bending band $\nu_2$. }}
\label{tunnelling}
\end{figure}

Of all the possible phosphine modes, the tunnelling effect should be
most prominent in the symmetric bending mode, $\nu_{2}$, as it is the
mode most strongly associated with the height of the pyramid formed
with the phosphorous atom on top. In ammonia the analogous
$\nu_{2}$ mode is known as the
inversion mode. Figure~\ref{tunnelling} shows schematically the
relationship between this mode and the barrier to tunnelling for the
phosphine molecule.

The ExoMol group works on constructing comprehensive line lists for
modelling the atmospheres of hot bodies such as cool stars and
exoplanets \cite{jt528}.  As part of this work we have computed two
line lists for $^{31}$PH$_3$ in its ground electronic
state.\cite{jt556,jt592} The more accurate of these line lists, called
SAlTY,\cite{jt592} contains 16 billion transitions between 9.8 million
energy levels and it is suitable for simulating spectra up to
temperatures of 1500~K. It covers wavenumbers up to 10~000 \cm\ and
includes all transitions to upper states with energies below $hc \cdot
18\,000$~cm$^{-1}$ and rotational excitation up to $J=46$.

The PH$_3$ line lists were computed by the variational solution of the Schr\"{o}dinger
equation for the rotation-vibration motion employing the
nuclear-motion program {\sc TROVE}.\cite{07YuThJe.method}  The
line lists were computed using C$_{\rm 3v}$(M) symmetry, considering phosphine
as a rigid molecule and thus with the potential barrier between the
two symmetry-equivalent minima effectively set to infinity.
Consequently, it originally neglected the possibility of a tunnelling mode.

Tunnelling can be considered a chemical reaction and as such it is
very sensitive to the shape of the potential energy surface (PES)
\cite{80Bell}. The SAlTY line list used a spectroscopically-refined
version of the {\it{ab initio}} (CCSD(T)/aug-cc-pV(Q+d)Z) potential
energy surface (PES).\cite{08OvThYu2.PH3} The value of splitting in
various vibrational states as well as the intensity of the
inversion-rotation and inversion-ro-vibrational lines can be computed
by adapting the procedure used to simulate the phosphine
spectrum to work with D$_{3h}$(M) symmetry. D$_{3h}$(M) is the permutation
inversion group for ammonia, since it is much less rigid molecule than phosphine.

{\sc TROVE} is used to compute differences between split energy levels.  Here we
employ the
same refined PES  as that used to compute the SAlTY line list
to predict the splitting between states of $A_1$ and $A_2$
symmetry for $J=0$, considering a zero point energy of 5232.26 \cm.
The potential energy function and the kinetic energy operator are expanded (six
and eight orders, respectively)
in terms of the
five non-linearized internal coordinates (three stretching and two deformational
bending) around
a symmetric one-dimensional non-rigid reference configuration represented by the
inversion mode.
The vibrational basis functions are obtained in a two-step contraction approach
as described by Yurchenko  {\it et al.}\cite{jt500} The stretching ($\nu_1$ and
$\nu_3$) primitive basis function $| v_{\rm str} \rangle$ ($v_{\rm str} =
0\ldots 7$) are obtained using the
Numerov-Cooley method.\cite{23Nuxxxx.method,61Coxxxx.method} Harmonic
oscillators are used as basis functions for the bending ($\nu_4$) primitive,  $| v_{\rm bend}, with
\rangle$ ($v_{\rm bend}=0\ldots 24$).
For the  $\nu_2$ inversion mode, primitive basis functions, $| v_{\rm inv} \rangle$, are used. These were also generated with the Numerov-Cooley method, with $v_{\rm inv} \leq 64$.

The phosphine barrier height values used in the TROVE input were
11~130.0 cm$^{-1}$ for the planar  local minimum with
P--H bonds at 1.3611~\AA, refined
from an {\it{ab initio}} value of 11353.6 cm$^{-1}$ for the local
minimum at 1.3858~\AA. Both the {\it{ab initio}} and refined
barrier heights are extrapolated values from the potential parameters
in the PES. These values are somewhat lower than the previous,
lower-level {\it ab initio} estimates but are in reasonable
agreement with the empirical estimate of Weston.\cite{54We}

To help assess the uncertainty in our predicted splittings,
calculations were made using two different PES surfaces, pre and post
refinement, corresponding to the surfaces used to calculate the
phosphine line list at room temperature,\cite{jt556} and the complete
SAlTY line list,\cite{jt592} respectively. Even though the refined PES
resulted in a significant improvement in the accuracy
of the overall phosphine
line list, the predicted splittings agreed completely up to four
significant figures.

Table \ref{split_size} shows
how the predicted splittings change as a function of  $\nu_{2}$
excitation.  Splittings in the
ground state are known to be extremely small\cite{92ScLaPy.PH3} (our calculations suggest about $10^{-10}$ \cm)
but increase significantly as the $\nu_{2}$ mode is excited.  All splitting predictions
are converged to
within 40~\%\ up to $7\nu_2$; use of even larger inversion basis sets ($v_{\rm inv}>64$) became numerically unstable.

\begin{table}[h!]
\caption{Calculated splitting for the ground state (GS), fundamental and excited bands of the bending mode $\nu_{2}$.}
.\label{split_size}
\small \tabcolsep=8pt
\renewcommand{\arraystretch}{1.4}
\centering
\begin{tabular}{{c}{l}{c}}
\hline\hline
{Band}& Energy\citep{jt592} & Splitting (cm$^{-1}$  \\
\hline

GS	& 0.000 &                     $  \leq 10^{-10}$	    \\
$\nu_{2}$&992.136	&         $ 2\times10^{-8}$	    \\
$2\nu_{2}$&	1972.576 & $6\times10^{-7}$		    \\
$3\nu_{2}$&	2940.810 & $1\times10^{-5}$	    \\
$4\nu_{2}$&	3895.685 &  0.0001	    	    \\
$5\nu_{2}$&	4837.223 &   0.0009	 	    \\
$6\nu_{2}$&	5767.008 &  0.0028			    \\
$7\nu_{2}$&	6687.601 &  0.0165	 	    \\
$8\nu_{2}$&	7598.124 &  0.0525			    \\
$9\nu_{2}$&	8494.683 &  0.6243			    \\
\hline\hline

\end{tabular}

\end{table}

Our tunnelling splitting for the overtones of $\nu_2$ are
somewhat larger than those predicted by previous
one-dimension studies.\cite{78SpPa,92ScLaPy.PH3} This could have been
anticipated as it is well-known\cite{96KlQuSu} that the treatment of
tunnelling which considers all dimensions of the problem leads to
increases in the magnitude of the splitting (or faster tunnelling).
Additionally, our lower value for the barrier height also contributes
to the larger predicted splitting values.

Any observation of the tunnelling splitting in PH$_3$ has to consider
a number of factors. First this splitting has to be distinguished from
the hyperfine structure.  The hyperfine splitting in PH$_3$ has been
observed\cite{13Muller.PH3, 06CaPu} to be less than 1 MHz or
4$\times$10$^{-5}$ cm$^{-1}$ and should not increase significantly
with vibrational excitation. Consequently, the splitting due to
inversion should be distinguishable from the hyperfine splitting for
all bands associated with vibrational excitation to $4\nu_{2}$ and
higher. Besides, the nuclear statistics of PH$_3$ as a D$_{3h}$(M) symmetry
molecule should be also taken into account. For example, as in the case of NH$_3$,
the ro-vibrational states of the $A_1'$ and $A_1''$ symmetries have zero
nuclear statistical weights $g_{\rm ns}$ and thus forbidden, with $g_{rm ns} = $ 8, 8, 4 and 4
for $A_2', A_2'', E'$ and $E''$, respectively.

\begin{figure}[h!]
\centering
\includegraphics[width=0.48\textwidth]{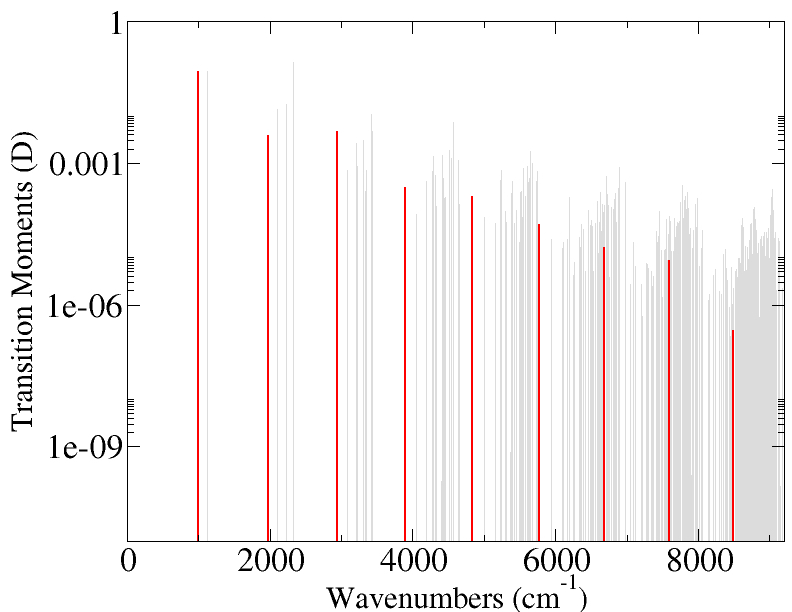}
\caption{\small{Transition moments (Debye, log scale) for excitations from the ground
vibrational states up to 16~000 \cm. Transitions associated
with $\nu_2$ mode excitations are highlighted.}}

\label{tms}
\end{figure}

 Due to their reasonably large energy splittings, promising regions of possible detection are those of
the symmetric bending bands, $\rm6\nu_2$ ($\approx$ 5800 cm$^{-1}$) and $\rm7\nu_2$ ($\approx$ 6700 cm$^{-1}$). Their splittings are predicted to be approximately 0.003 cm$^{-1}$ and 0.02 cm$^{-1}$ for $\rm6\nu_2$ and $\rm7\nu_2$, respectively.
Our calculations suggest that most intense lines in this band have
intensities of about 10$^{-24}$ cm/molecule at $T=296$~K and should be
easily observable with modern instruments.  Figure \ref{tms}
summarises the dipole transition moments to various vibrational states
from the vibrational ground state; transitions to the $\nu_2$ overtone
series associated with the tunnelling motion are highlighted.

\begin{figure}
\centering

\includegraphics[width=0.4\textwidth]{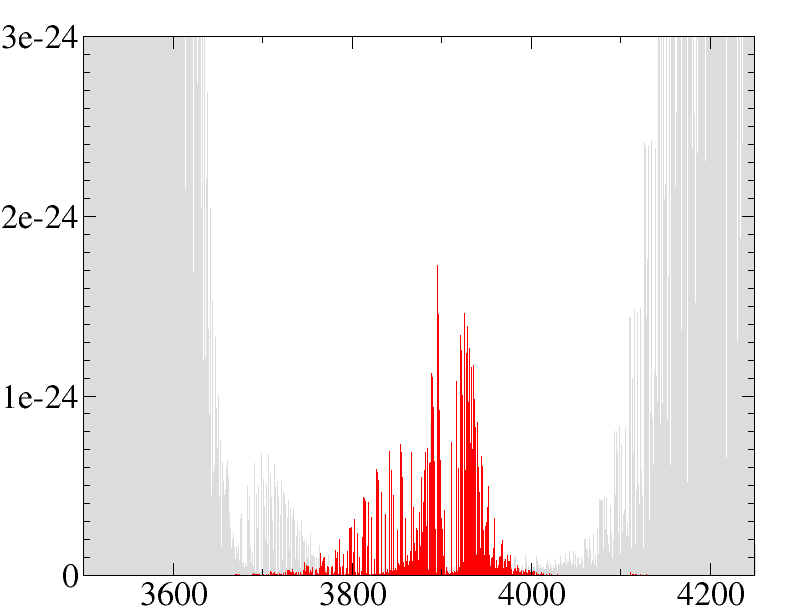}
 \includegraphics[width=0.4\textwidth]{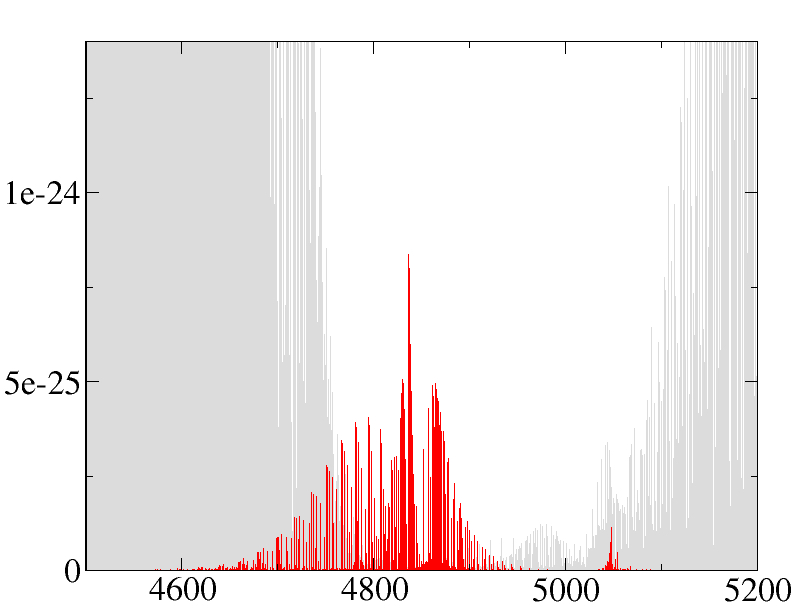}
 \includegraphics[width=0.4\textwidth]{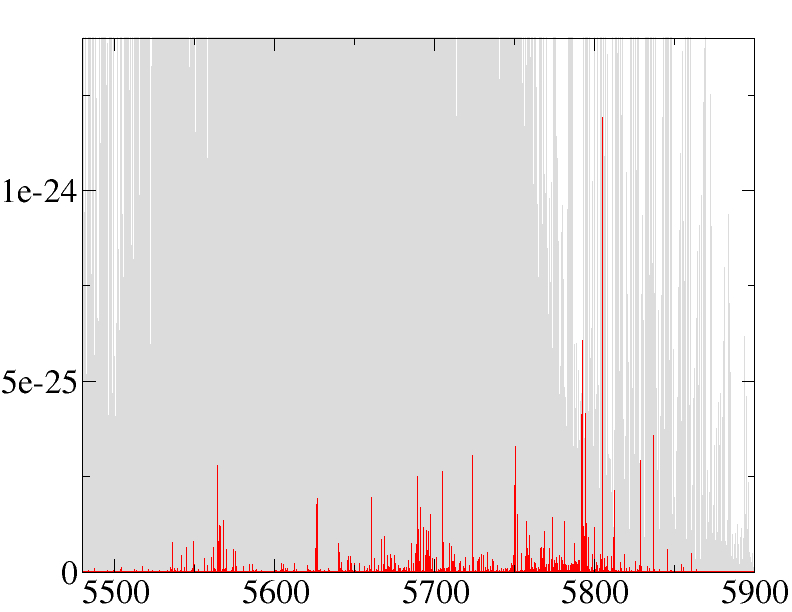}
 \includegraphics[width=0.4\textwidth]{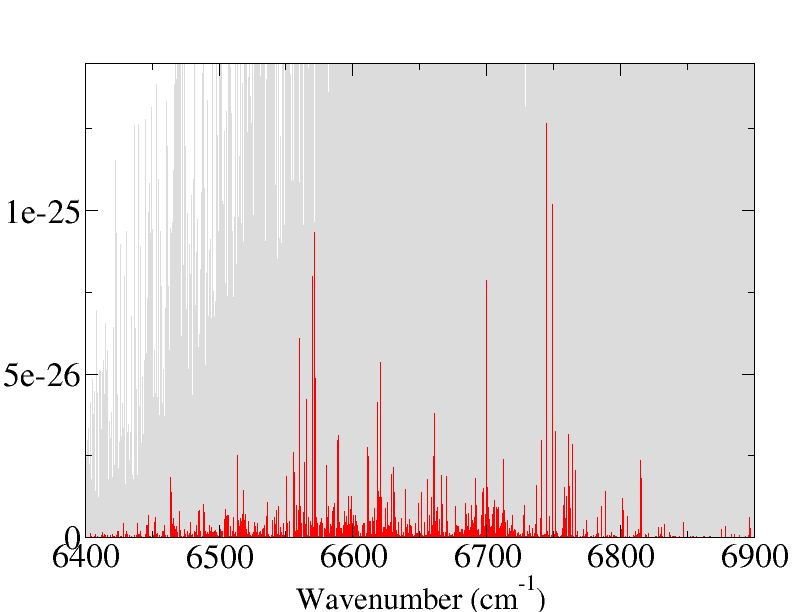}

\caption{Contrast between the (top to bottom) $\rm4\nu_2$, $\rm5\nu_2$, $\rm6\nu_2$ and $\rm7\nu_2$ bands and the neighbouring transitions at $T=296$~K.
The SAlTY absorption intensities (cm/molecule)  are computed using a C$_{\rm 3v}$(M) model with PH$_3$ as a rigid molecule, i.e.
neglecting the inversion splitting.\protect\cite{jt592} }
\label{contrasts}
\end{figure}

However, detection
will also depend on the location of the splitting transitions as it
may be difficult to distinguish the $\nu_2$ bands in regions of the
spectrum that are strongly populated by other bands.
Figure \ref{contrasts} highlights the spectroscopic regions where splitting
could be detected in the
context of the surrounding spectrum for the $4\nu_2$, $5\nu_2$,
$6\nu_2$ and $7\nu_2$ bands. In this context, the positions of the $4\nu_2$ and $5\nu_2$ bands appear to be particularly promising for investigation, since they can be mostly
isolated from the surrounding stronger bands.

\begin{figure}[h!]
\centering

\includegraphics[width=0.45\textwidth]{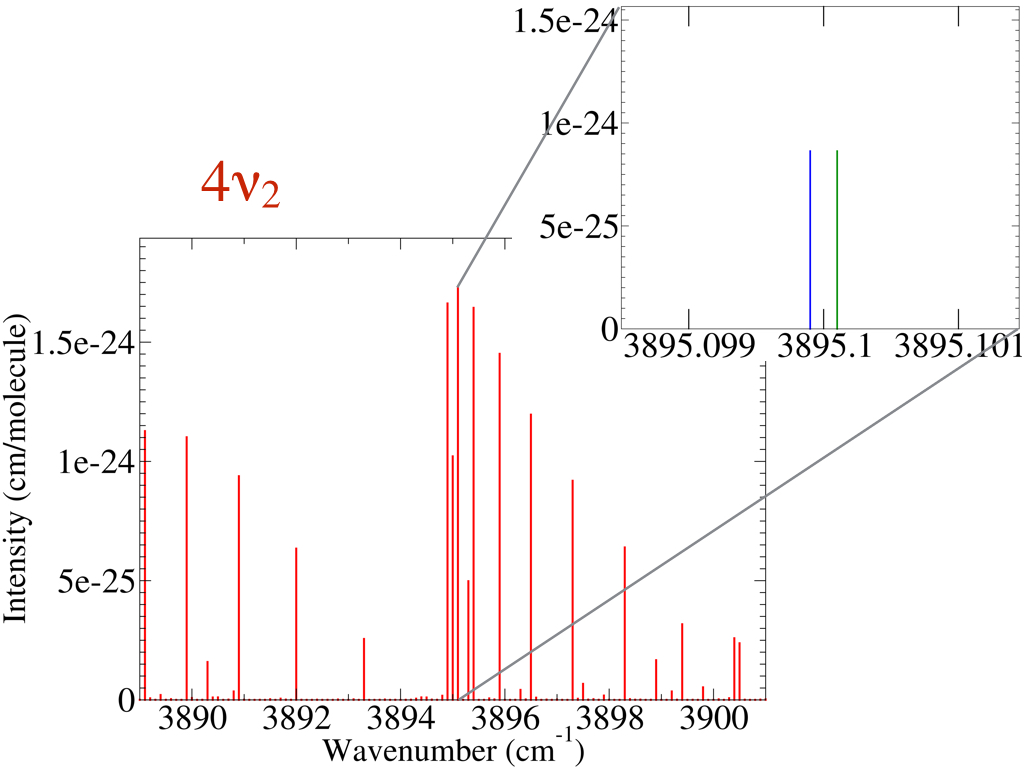}
 \includegraphics[width=0.45\textwidth]{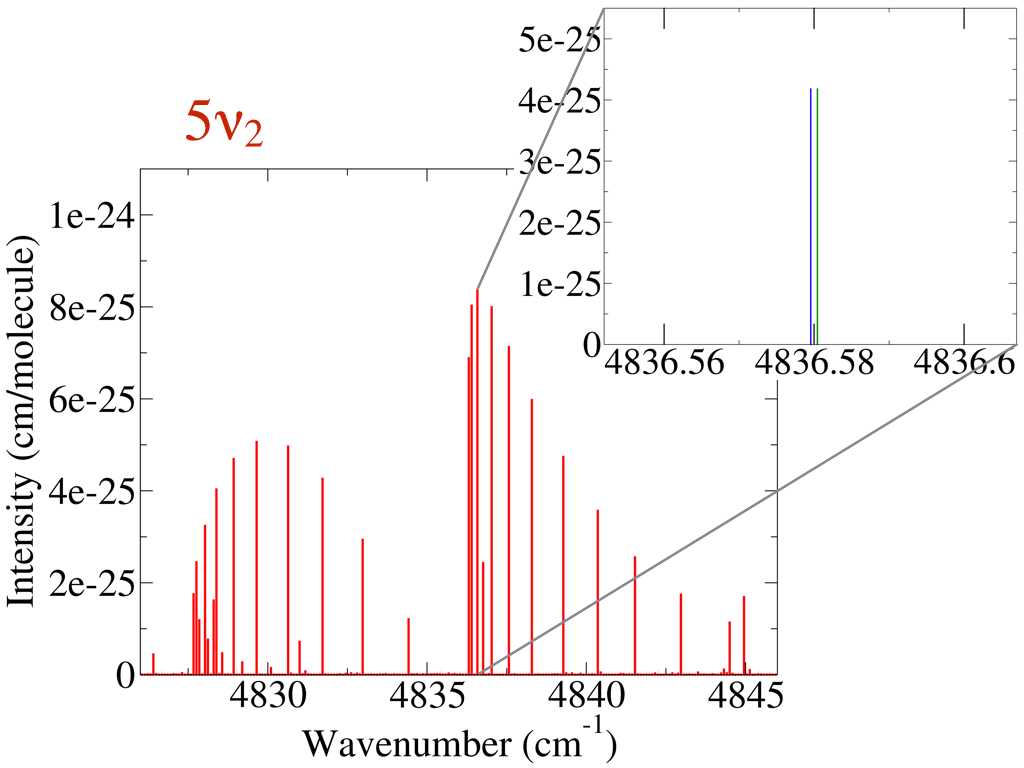}

 \caption{Comparison between predicted phosphine spectra without
   (red) and with (blue and green) the inclusion of tunnelling motion, for
   the strongest transitions in the $\rm4\nu_2$ and the $\rm5\nu_2$ overtone bands. The ro-vibrational splitting is estimated using the pure vibrational values
   from Table \ref{split_size}. The SAlTY line list is used to simulate absorption intensities for
   a temperature of 296 K.}

\label{splits_1}
\end{figure}

\begin{figure}[h!]
\centering

 \includegraphics[width=0.45\textwidth]{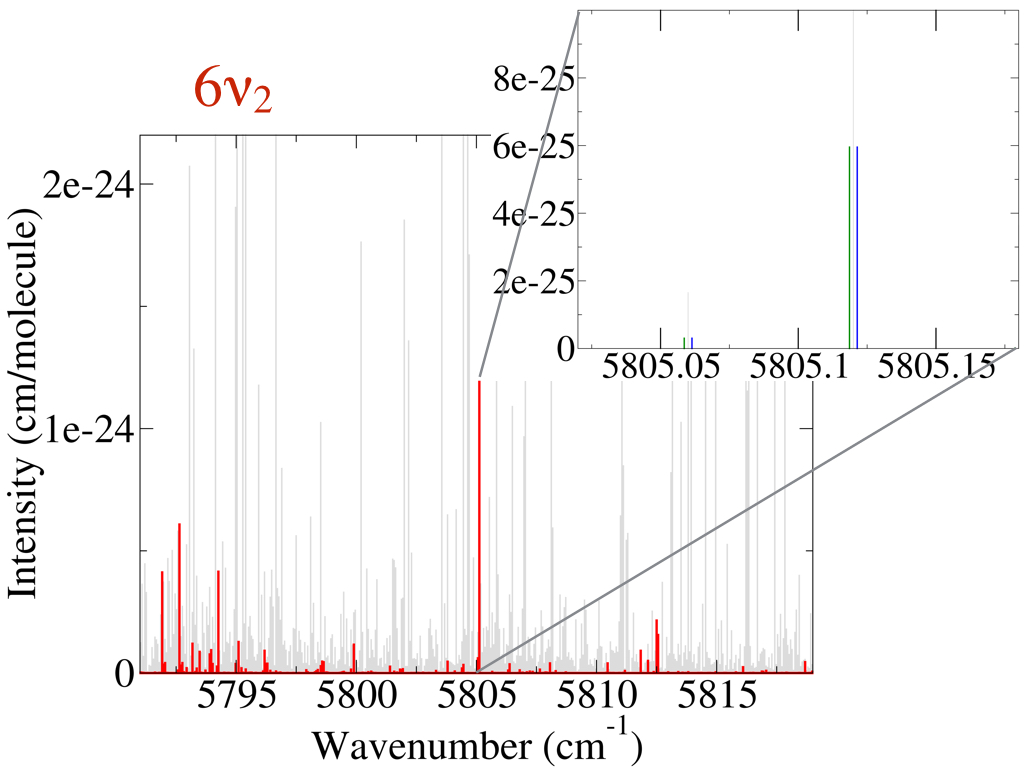}
 \includegraphics[width=0.45\textwidth]{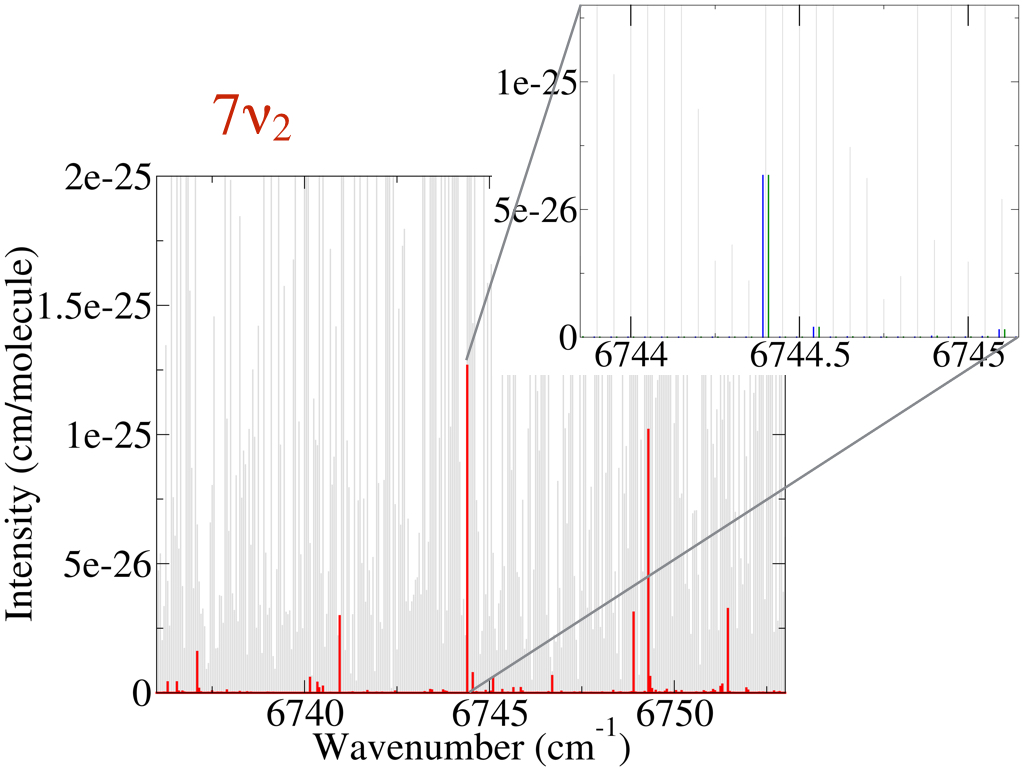}
 \caption{Comparison between predicted phosphine spectra without
   (red) and with (blue and green) the inclusion of tunnelling motion, for
   the strongest transitions in the $\rm6\nu_2$ and
   $\rm7\nu_2$ overtone bands. The ro-vibrational splitting is estimated using the pure vibrational values
   from Table \ref{split_size}. The SAlTY line list is used to simulate absorption intensities for
   a temperature of 296 K. }

\label{splits_2}
\end{figure}

 Figure \ref{splits_1} shows the predicted spectra in the region of the strongest transitions for $4\nu_2$ and $5\nu_2$ bands, comparing spectra when the molecule
is allowed to undergo inversion and when tunnelling is not permitted. Additionally, Figure \ref{splits_2} shows how the $6\nu_2$ and $7\nu_2$ bands will be harder to detect amongst the surrounding bands, despite having much larger splitting values.

\vspace{0.2in}

Our calculations show that the $\nu_2$ overtones display
splittings of a magnitude that should be resolvable with modern
experiments. We therefore hope that
the theoretical predictions of phosphine tunnelling
shown here will be validated with experimental detection in the near
future. Simulated spectra for other regions and/or conditions
can be provided by the authors to aid this process.

\begin{acknowledgments}
  This work is supported by ERC Advanced Investigator Project 267219.
  We would like to thank Laura McKemmish, Ahmed Al-Refaie, Jack D. Franklin and William Azubuike for their
  support and advice.

\end{acknowledgments}

\vspace{0.3in}


\begin{thebibliography}{26}%
\makeatletter
\providecommand \@ifxundefined [1]{%
 \@ifx{#1\undefined}
}%
\providecommand \@ifnum [1]{%
 \ifnum #1\expandafter \@firstoftwo
 \else \expandafter \@secondoftwo
 \fi
}%
\providecommand \@ifx [1]{%
 \ifx #1\expandafter \@firstoftwo
 \else \expandafter \@secondoftwo
 \fi
}%
\providecommand \natexlab [1]{#1}%
\providecommand \enquote  [1]{``#1''}%
\providecommand \bibnamefont  [1]{#1}%
\providecommand \bibfnamefont [1]{#1}%
\providecommand \citenamefont [1]{#1}%
\providecommand \href@noop [0]{\@secondoftwo}%
\providecommand \href [0]{\begingroup \@sanitize@url \@href}%
\providecommand \@href[1]{\@@startlink{#1}\@@href}%
\providecommand \@@href[1]{\endgroup#1\@@endlink}%
\providecommand \@sanitize@url [0]{\catcode `\\12\catcode `\$12\catcode
  `\&12\catcode `\#12\catcode `\^12\catcode `\_12\catcode `\%12\relax}%
\providecommand \@@startlink[1]{}%
\providecommand \@@endlink[0]{}%
\providecommand \url  [0]{\begingroup\@sanitize@url \@url }%
\providecommand \@url [1]{\endgroup\@href {#1}{\urlprefix }}%
\providecommand \urlprefix  [0]{URL }%
\providecommand \Eprint [0]{\href }%
\providecommand \doibase [0]{http://dx.doi.org/}%
\providecommand \selectlanguage [0]{\@gobble}%
\providecommand \bibinfo  [0]{\@secondoftwo}%
\providecommand \bibfield  [0]{\@secondoftwo}%
\providecommand \translation [1]{[#1]}%
\providecommand \BibitemOpen [0]{}%
\providecommand \bibitemStop [0]{}%
\providecommand \bibitemNoStop [0]{.\EOS\space}%
\providecommand \EOS [0]{\spacefactor3000\relax}%
\providecommand \BibitemShut  [1]{\csname bibitem#1\endcsname}%
\let\auto@bib@innerbib\@empty
\bibitem [{\citenamefont {Bell}(1980)}]{80Bell}%
  \BibitemOpen
  \bibfield  {author} {\bibinfo {author} {\bibfnamefont {R.~P.}\ \bibnamefont
  {Bell}},\ }\href@noop {} {\emph {\bibinfo {title} {The tunnel effect in
  chemistry}}}\ (\bibinfo  {publisher} {Springer},\ \bibinfo {year} {1980})\
  pp.\ \bibinfo {pages} {51--76}\BibitemShut {NoStop}%
\bibitem [{\citenamefont {Dennison}\ and\ \citenamefont
  {Uhlenbeck}(1932)}]{32DeUh}%
  \BibitemOpen
  \bibfield  {author} {\bibinfo {author} {\bibfnamefont {D.~M.}\ \bibnamefont
  {Dennison}}\ and\ \bibinfo {author} {\bibfnamefont {G.~E.}\ \bibnamefont
  {Uhlenbeck}},\ }\href@noop {} {\bibfield  {journal} {\bibinfo  {journal}
  {Phys. Rev.}\ }\textbf {\bibinfo {volume} {41}},\ \bibinfo {pages} {313}
  (\bibinfo {year} {1932})}\BibitemShut {NoStop}%
\bibitem [{\citenamefont {Cleeton}\ and\ \citenamefont
  {Williams}(1934)}]{34ClWiXX}%
  \BibitemOpen
  \bibfield  {author} {\bibinfo {author} {\bibfnamefont {C.~E.}\ \bibnamefont
  {Cleeton}}\ and\ \bibinfo {author} {\bibfnamefont {N.~H.}\ \bibnamefont
  {Williams}},\ }\href@noop {} {\bibfield  {journal} {\bibinfo  {journal}
  {Phys. Rev.}\ }\textbf {\bibinfo {volume} {45}},\ \bibinfo {pages} {234}
  (\bibinfo {year} {1934})}\BibitemShut {NoStop}%
\bibitem [{\citenamefont {Csaszar}\ and\ \citenamefont
  {Furtenbacher}(2016)}]{16CsFuxx.NH3}%
  \BibitemOpen
  \bibfield  {author} {\bibinfo {author} {\bibfnamefont {A.~G.}\ \bibnamefont
  {Csaszar}}\ and\ \bibinfo {author} {\bibfnamefont {T.}~\bibnamefont
  {Furtenbacher}},\ }\href {\doibase {10.1039/c5cp04270d}} {\bibfield
  {journal} {\bibinfo  {journal} {Phys. Chem. Chem. Phys.}\ }\textbf {\bibinfo
  {volume} {{18}}},\ \bibinfo {pages} {1092} (\bibinfo {year}
  {{2016}})}\BibitemShut {NoStop}%
\bibitem [{\citenamefont {Wright}\ and\ \citenamefont
  {Randall}(1933)}]{33WrRa}%
  \BibitemOpen
  \bibfield  {author} {\bibinfo {author} {\bibfnamefont {N.}~\bibnamefont
  {Wright}}\ and\ \bibinfo {author} {\bibfnamefont {H.~M.}\ \bibnamefont
  {Randall}},\ }\href@noop {} {\bibfield  {journal} {\bibinfo  {journal} {Phys.
  Rev.}\ }\textbf {\bibinfo {volume} {44}},\ \bibinfo {pages} {391} (\bibinfo
  {year} {1933})}\BibitemShut {NoStop}%
\bibitem [{\citenamefont {Stroup}\ \emph {et~al.}(1953)\citenamefont {Stroup},
  \citenamefont {Oetjen},\ and\ \citenamefont {Bell}}]{53StOeBe}%
  \BibitemOpen
  \bibfield  {author} {\bibinfo {author} {\bibfnamefont {R.~E.}\ \bibnamefont
  {Stroup}}, \bibinfo {author} {\bibfnamefont {R.~A.}\ \bibnamefont {Oetjen}},
  \ and\ \bibinfo {author} {\bibfnamefont {E.~E.}\ \bibnamefont {Bell}},\
  }\href@noop {} {\bibfield  {journal} {\bibinfo  {journal} {J. Opt. Soc. Am.}\
  }\textbf {\bibinfo {volume} {43}},\ \bibinfo {pages} {1096} (\bibinfo {year}
  {1953})}\BibitemShut {NoStop}%
\bibitem [{\citenamefont {Helminger}\ and\ \citenamefont
  {Gordy}(1969)}]{69HeGo}%
  \BibitemOpen
  \bibfield  {author} {\bibinfo {author} {\bibfnamefont {P.}~\bibnamefont
  {Helminger}}\ and\ \bibinfo {author} {\bibfnamefont {W.}~\bibnamefont
  {Gordy}},\ }\href@noop {} {\bibfield  {journal} {\bibinfo  {journal} {Phys.
  Rev.}\ }\textbf {\bibinfo {volume} {188}},\ \bibinfo {pages} {100} (\bibinfo
  {year} {1969})}\BibitemShut {NoStop}%
\bibitem [{\citenamefont {Davies}\ \emph {et~al.}(1971)\citenamefont {Davies},
  \citenamefont {Neumann}, \citenamefont {Wofsy},\ and\ \citenamefont
  {Klemperer}}]{71DaNeWo}%
  \BibitemOpen
  \bibfield  {author} {\bibinfo {author} {\bibfnamefont {P.~B.}\ \bibnamefont
  {Davies}}, \bibinfo {author} {\bibfnamefont {R.~M.}\ \bibnamefont {Neumann}},
  \bibinfo {author} {\bibfnamefont {S.~C.}\ \bibnamefont {Wofsy}}, \ and\
  \bibinfo {author} {\bibfnamefont {W.}~\bibnamefont {Klemperer}},\ }\href@noop
  {} {\bibfield  {journal} {\bibinfo  {journal} {J. Chem. Phys.}\ }\textbf
  {\bibinfo {volume} {55}},\ \bibinfo {pages} {3564} (\bibinfo {year}
  {1971})}\BibitemShut {NoStop}%
\bibitem [{\citenamefont {Maki}\ \emph {et~al.}(1973)\citenamefont {Maki},
  \citenamefont {Sams},\ and\ \citenamefont {Olson}}]{73MaSaOl}%
  \BibitemOpen
  \bibfield  {author} {\bibinfo {author} {\bibfnamefont {A.~G.}\ \bibnamefont
  {Maki}}, \bibinfo {author} {\bibfnamefont {R.~L.}\ \bibnamefont {Sams}}, \
  and\ \bibinfo {author} {\bibfnamefont {W.~B.}\ \bibnamefont {Olson}},\
  }\href@noop {} {\bibfield  {journal} {\bibinfo  {journal} {J. Chem. Phys.}\
  }\textbf {\bibinfo {volume} {58}},\ \bibinfo {pages} {4502} (\bibinfo {year}
  {1973})}\BibitemShut {NoStop}%
\bibitem [{\citenamefont {{\v S}pirko}\ and\ \citenamefont {Papou{\v
  s}ek}(1978)}]{78SpPa}%
  \BibitemOpen
  \bibfield  {author} {\bibinfo {author} {\bibfnamefont {V.}~\bibnamefont {{\v
  S}pirko}}\ and\ \bibinfo {author} {\bibfnamefont {D.}~\bibnamefont {Papou{\v
  s}ek}},\ }\href {\doibase 10.1080/00268977800101941} {\bibfield  {journal}
  {\bibinfo  {journal} {Mol. Phys.}\ }\textbf {\bibinfo {volume} {36}},\
  \bibinfo {pages} {791} (\bibinfo {year} {1978})}\BibitemShut {NoStop}%
\bibitem [{\citenamefont {Belov}\ \emph {et~al.}(1981)\citenamefont {Belov},
  \citenamefont {Burenin}, \citenamefont {Gershtein}, \citenamefont {Krupnov},
  \citenamefont {Markov}, \citenamefont {Maslovsky},\ and\ \citenamefont
  {Shapin}}]{81BeBuGe}%
  \BibitemOpen
  \bibfield  {author} {\bibinfo {author} {\bibfnamefont {S.~P.}\ \bibnamefont
  {Belov}}, \bibinfo {author} {\bibfnamefont {A.~V.}\ \bibnamefont {Burenin}},
  \bibinfo {author} {\bibfnamefont {L.~I.}\ \bibnamefont {Gershtein}}, \bibinfo
  {author} {\bibfnamefont {A.~F.}\ \bibnamefont {Krupnov}}, \bibinfo {author}
  {\bibfnamefont {V.~N.}\ \bibnamefont {Markov}}, \bibinfo {author}
  {\bibfnamefont {A.~V.}\ \bibnamefont {Maslovsky}}, \ and\ \bibinfo {author}
  {\bibfnamefont {S.~M.}\ \bibnamefont {Shapin}},\ }\href@noop {} {\bibfield
  {journal} {\bibinfo  {journal} {J. Mol. Spectrosc.}\ }\textbf {\bibinfo
  {volume} {86}},\ \bibinfo {pages} {184} (\bibinfo {year} {1981})}\BibitemShut
  {NoStop}%
\bibitem [{\citenamefont {Schwerdtfeger}\ \emph {et~al.}(1992)\citenamefont
  {Schwerdtfeger}, \citenamefont {Laakkonen},\ and\ \citenamefont
  {Pyykk{\"o}}}]{92ScLaPy.PH3}%
  \BibitemOpen
  \bibfield  {author} {\bibinfo {author} {\bibfnamefont {P.}~\bibnamefont
  {Schwerdtfeger}}, \bibinfo {author} {\bibfnamefont {L.~J.}\ \bibnamefont
  {Laakkonen}}, \ and\ \bibinfo {author} {\bibfnamefont {P.}~\bibnamefont
  {Pyykk{\"o}}},\ }\href@noop {} {\bibfield  {journal} {\bibinfo  {journal} {J.
  Chem. Phys.}\ }\textbf {\bibinfo {volume} {96}},\ \bibinfo {pages} {6807}
  (\bibinfo {year} {1992})}\BibitemShut {NoStop}%
\bibitem [{\citenamefont {Weston~Jr}(1954)}]{54We}%
  \BibitemOpen
  \bibfield  {author} {\bibinfo {author} {\bibfnamefont {R.~E.}\ \bibnamefont
  {Weston~Jr}},\ }\href@noop {} {\bibfield  {journal} {\bibinfo  {journal} {J.
  Am. Chem. Soc.}\ }\textbf {\bibinfo {volume} {76}},\ \bibinfo {pages} {2645}
  (\bibinfo {year} {1954})}\BibitemShut {NoStop}%
\bibitem [{\citenamefont {Sousa-Silva}\ \emph {et~al.}(2013)\citenamefont
  {Sousa-Silva}, \citenamefont {Yurchenko},\ and\ \citenamefont
  {Tennyson}}]{jt556}%
  \BibitemOpen
  \bibfield  {author} {\bibinfo {author} {\bibfnamefont {C.}~\bibnamefont
  {Sousa-Silva}}, \bibinfo {author} {\bibfnamefont {S.~N.}\ \bibnamefont
  {Yurchenko}}, \ and\ \bibinfo {author} {\bibfnamefont {J.}~\bibnamefont
  {Tennyson}},\ }\href@noop {} {\bibfield  {journal} {\bibinfo  {journal} {J.
  Mol. Spectrosc.}\ }\textbf {\bibinfo {volume} {288}},\ \bibinfo {pages} {28}
  (\bibinfo {year} {2013})}\BibitemShut {NoStop}%
\bibitem [{\citenamefont {Malathy~Devi}\ \emph {et~al.}(2014)\citenamefont
  {Malathy~Devi}, \citenamefont {Kleiner}, \citenamefont {Sams}, \citenamefont
  {Brown}, \citenamefont {Benner},\ and\ \citenamefont
  {Fletcher}}]{14DeKlSa.PH3}%
  \BibitemOpen
  \bibfield  {author} {\bibinfo {author} {\bibfnamefont {V.}~\bibnamefont
  {Malathy~Devi}}, \bibinfo {author} {\bibfnamefont {I.}~\bibnamefont
  {Kleiner}}, \bibinfo {author} {\bibfnamefont {R.~L.}\ \bibnamefont {Sams}},
  \bibinfo {author} {\bibfnamefont {L.~R.}\ \bibnamefont {Brown}}, \bibinfo
  {author} {\bibfnamefont {D.~C.}\ \bibnamefont {Benner}}, \ and\ \bibinfo
  {author} {\bibfnamefont {L.~N.}\ \bibnamefont {Fletcher}},\ }\href@noop {}
  {\bibfield  {journal} {\bibinfo  {journal} {J. Mol. Spectrosc.}\ }\textbf
  {\bibinfo {volume} {298}},\ \bibinfo {pages} {11} (\bibinfo {year}
  {2014})}\BibitemShut {NoStop}%
\bibitem [{\citenamefont {Ulenikov}\ \emph {et~al.}(2002)\citenamefont
  {Ulenikov}, \citenamefont {Bekhtereva}, \citenamefont {Kozinskaia},
  \citenamefont {Zheng}, \citenamefont {He}, \citenamefont {Hu}, \citenamefont
  {Zhu}, \citenamefont {Leroy},\ and\ \citenamefont {Pluchart}}]{02UlBeKo.PH3}%
  \BibitemOpen
  \bibfield  {author} {\bibinfo {author} {\bibfnamefont {O.~N.}\ \bibnamefont
  {Ulenikov}}, \bibinfo {author} {\bibfnamefont {E.~S.}\ \bibnamefont
  {Bekhtereva}}, \bibinfo {author} {\bibfnamefont {V.~A.}\ \bibnamefont
  {Kozinskaia}}, \bibinfo {author} {\bibfnamefont {J.~J.}\ \bibnamefont
  {Zheng}}, \bibinfo {author} {\bibfnamefont {S.~G.}\ \bibnamefont {He}},
  \bibinfo {author} {\bibfnamefont {S.~M.}\ \bibnamefont {Hu}}, \bibinfo
  {author} {\bibfnamefont {Q.~S.}\ \bibnamefont {Zhu}}, \bibinfo {author}
  {\bibfnamefont {C.}~\bibnamefont {Leroy}}, \ and\ \bibinfo {author}
  {\bibfnamefont {L.}~\bibnamefont {Pluchart}},\ }\href@noop {} {\bibfield
  {journal} {\bibinfo  {journal} {J. Mol. Spectrosc.}\ }\textbf {\bibinfo
  {volume} {215}},\ \bibinfo {pages} {295} (\bibinfo {year}
  {2002})}\BibitemShut {NoStop}%
\bibitem [{\citenamefont {Tennyson}\ and\ \citenamefont
  {Yurchenko}(2012)}]{jt528}%
  \BibitemOpen
  \bibfield  {author} {\bibinfo {author} {\bibfnamefont {J.}~\bibnamefont
  {Tennyson}}\ and\ \bibinfo {author} {\bibfnamefont {S.~N.}\ \bibnamefont
  {Yurchenko}},\ }\href@noop {} {\bibfield  {journal} {\bibinfo  {journal}
  {Mon. Not. R. Astron. Soc.}\ }\textbf {\bibinfo {volume} {425}},\ \bibinfo
  {pages} {21} (\bibinfo {year} {2012})}\BibitemShut {NoStop}%
\bibitem [{\citenamefont {Sousa-Silva}\ \emph {et~al.}(2015)\citenamefont
  {Sousa-Silva}, \citenamefont {Al-Refaie}, \citenamefont {Tennyson},\ and\
  \citenamefont {Yurchenko}}]{jt592}%
  \BibitemOpen
  \bibfield  {author} {\bibinfo {author} {\bibfnamefont {C.}~\bibnamefont
  {Sousa-Silva}}, \bibinfo {author} {\bibfnamefont {A.~F.}\ \bibnamefont
  {Al-Refaie}}, \bibinfo {author} {\bibfnamefont {J.}~\bibnamefont {Tennyson}},
  \ and\ \bibinfo {author} {\bibfnamefont {S.~N.}\ \bibnamefont {Yurchenko}},\
  }\href {\doibase 10.1093/mnras/stu2246} {\bibfield  {journal} {\bibinfo
  {journal} {Mon. Not. R. Astron. Soc.}\ }\textbf {\bibinfo {volume} {446}},\
  \bibinfo {pages} {2337} (\bibinfo {year} {2015})}\BibitemShut {NoStop}%
\bibitem [{\citenamefont {Yurchenko}\ \emph {et~al.}(2007)\citenamefont
  {Yurchenko}, \citenamefont {Thiel},\ and\ \citenamefont
  {Jensen}}]{07YuThJe.method}%
  \BibitemOpen
  \bibfield  {author} {\bibinfo {author} {\bibfnamefont {S.~N.}\ \bibnamefont
  {Yurchenko}}, \bibinfo {author} {\bibfnamefont {W.}~\bibnamefont {Thiel}}, \
  and\ \bibinfo {author} {\bibfnamefont {P.}~\bibnamefont {Jensen}},\ }\href
  {\doibase 10.1016/j.jms.2007.07.009} {\bibfield  {journal} {\bibinfo
  {journal} {J. Mol. Spectrosc.}\ }\textbf {\bibinfo {volume} {245}},\ \bibinfo
  {pages} {126} (\bibinfo {year} {2007})}\BibitemShut {NoStop}%
\bibitem [{\citenamefont {Ovsyannikov}\ \emph {et~al.}(2008)\citenamefont
  {Ovsyannikov}, \citenamefont {Thiel}, \citenamefont {Yurchenko},
  \citenamefont {Carvajal},\ and\ \citenamefont {Jensen}}]{08OvThYu2.PH3}%
  \BibitemOpen
  \bibfield  {author} {\bibinfo {author} {\bibfnamefont {R.~I.}\ \bibnamefont
  {Ovsyannikov}}, \bibinfo {author} {\bibfnamefont {W.}~\bibnamefont {Thiel}},
  \bibinfo {author} {\bibfnamefont {S.~N.}\ \bibnamefont {Yurchenko}}, \bibinfo
  {author} {\bibfnamefont {M.}~\bibnamefont {Carvajal}}, \ and\ \bibinfo
  {author} {\bibfnamefont {P.}~\bibnamefont {Jensen}},\ }\href@noop {}
  {\bibfield  {journal} {\bibinfo  {journal} {J. Chem. Phys.}\ }\textbf
  {\bibinfo {volume} {129}},\ \bibinfo {pages} {044309} (\bibinfo {year}
  {2008})}\BibitemShut {NoStop}%
\bibitem [{\citenamefont {Yurchenko}\ \emph {et~al.}(2011)\citenamefont
  {Yurchenko}, \citenamefont {Barber},\ and\ \citenamefont {Tennyson}}]{jt500}%
  \BibitemOpen
  \bibfield  {author} {\bibinfo {author} {\bibfnamefont {S.~N.}\ \bibnamefont
  {Yurchenko}}, \bibinfo {author} {\bibfnamefont {R.~J.}\ \bibnamefont
  {Barber}}, \ and\ \bibinfo {author} {\bibfnamefont {J.}~\bibnamefont
  {Tennyson}},\ }\href@noop {} {\bibfield  {journal} {\bibinfo  {journal} {Mon.
  Not. R. Astron. Soc.}\ }\textbf {\bibinfo {volume} {413}},\ \bibinfo {pages}
  {1828} (\bibinfo {year} {2011})}\BibitemShut {NoStop}%
\bibitem [{\citenamefont {Noumeroff}(1923)}]{23Nuxxxx.method}%
  \BibitemOpen
  \bibfield  {author} {\bibinfo {author} {\bibfnamefont {B.}~\bibnamefont
  {Noumeroff}},\ }\enquote {\bibinfo {title} {M\'{e}thode nouvelle de la
  d{\'e}termination des orbites et le calcul des \'{e}́ph\'{e}m\'{e}́rides en
  tenant compte des perturbations},}\ in\ \href@noop {} {\emph {\bibinfo
  {booktitle} {Trudy Glavnoi Rossiiskoi Astrofizicheskoj Observatorii}}},\
  Vol.~\bibinfo {volume} {2}\ (\bibinfo  {publisher} {Moscow, Gosudarsvennoe
  Izdatel'stvo},\ \bibinfo {year} {1923})\ pp.\ \bibinfo {pages}
  {188--259}\BibitemShut {NoStop}%
\bibitem [{\citenamefont {Cooley}(1961)}]{61Coxxxx.method}%
  \BibitemOpen
  \bibfield  {author} {\bibinfo {author} {\bibfnamefont {J.~W.}\ \bibnamefont
  {Cooley}},\ }\href {\doibase
  http://dx.doi.org/10.1090/S0025-5718-1961-0129566-X} {\bibfield  {journal}
  {\bibinfo  {journal} {Math. Comp.}\ }\textbf {\bibinfo {volume} {15}},\
  \bibinfo {pages} {363} (\bibinfo {year} {1961})}\BibitemShut {NoStop}%
\bibitem [{\citenamefont {Klopper}\ \emph {et~al.}(1996)\citenamefont
  {Klopper}, \citenamefont {Quack},\ and\ \citenamefont {Suhm}}]{96KlQuSu}%
  \BibitemOpen
  \bibfield  {author} {\bibinfo {author} {\bibfnamefont {W.}~\bibnamefont
  {Klopper}}, \bibinfo {author} {\bibfnamefont {M.}~\bibnamefont {Quack}}, \
  and\ \bibinfo {author} {\bibfnamefont {M.~A.}\ \bibnamefont {Suhm}},\ }\href
  {\doibase 10.1016/0009-2614(96)00901-3} {\bibfield  {journal} {\bibinfo
  {journal} {Chem. Phys. Lett.}\ }\textbf {\bibinfo {volume} {{261}}},\
  \bibinfo {pages} {35} (\bibinfo {year} {{1996}})}\BibitemShut {NoStop}%
\bibitem [{\citenamefont {M{\"u}ller}(2013)}]{13Muller.PH3}%
  \BibitemOpen
  \bibfield  {author} {\bibinfo {author} {\bibfnamefont {H.~S.~P.}\
  \bibnamefont {M{\"u}ller}},\ }\href@noop {} {\bibfield  {journal} {\bibinfo
  {journal} {J. Quant. Spectrosc. Radiat. Transf.}\ }\textbf {\bibinfo {volume}
  {130}},\ \bibinfo {pages} {335} (\bibinfo {year} {2013})}\BibitemShut
  {NoStop}%
\bibitem [{\citenamefont {Cazzoli}\ and\ \citenamefont
  {Puzzarini}(2006)}]{06CaPu}%
  \BibitemOpen
  \bibfield  {author} {\bibinfo {author} {\bibfnamefont {G.}~\bibnamefont
  {Cazzoli}}\ and\ \bibinfo {author} {\bibfnamefont {C.}~\bibnamefont
  {Puzzarini}},\ }\href@noop {} {\bibfield  {journal} {\bibinfo  {journal} {J.
  Mol. Spectrosc.}\ }\textbf {\bibinfo {volume} {239}},\ \bibinfo {pages} {64}
  (\bibinfo {year} {2006})}\BibitemShut {NoStop}%
\end{thebibliography}

%

\end{document}